\documentclass{ieeeaccess}
\usepackage{algorithmic}
\usepackage{algorithm}
\usepackage{array}
\usepackage{textcomp}
\usepackage{stfloats}
\usepackage{url}
\usepackage{cite}
\usepackage{verbatim}
\usepackage[pdftex]{graphicx}
\usepackage{color}
\usepackage{transparent}
\usepackage{import}
\usepackage{subfigure}
\usepackage{balance}
\usepackage{multirow}
\DeclareGraphicsExtensions{.eps}
\usepackage{epstopdf}
\usepackage{epsfig}
\usepackage{gensymb}
\usepackage{soul}
\usepackage{mathrsfs}
\usepackage{float}
\usepackage{amsmath,amssymb,amsfonts}
\usepackage{cleveref}

\usepackage{bm}
\makeatletter
\AtBeginDocument{\DeclareMathVersion{bold}
\SetSymbolFont{operators}{bold}{T1}{times}{b}{n}
\SetSymbolFont{NewLetters}{bold}{T1}{times}{b}{it}
\SetMathAlphabet{\mathrm}{bold}{T1}{times}{b}{n}
\SetMathAlphabet{\mathit}{bold}{T1}{times}{b}{it}
\SetMathAlphabet{\mathbf}{bold}{T1}{times}{b}{n}
\SetMathAlphabet{\mathtt}{bold}{OT1}{pcr}{b}{n}
\SetSymbolFont{symbols}{bold}{OMS}{cmsy}{b}{n}
\renewcommand\boldmath{\@nomath\boldmath\mathversion{bold}}}
\makeatother

\def\BibTeX{{\rm B\kern-.05em{\sc i\kern-.025em b}\kern-.08em
    T\kern-.1667em\lower.7ex\hbox{E}\kern-.125emX}}

\begin{document}
\history{Date of publication xxxx 00, 0000, date of current version xxxx 00, 0000.}
\doi{10.1109/ACCESS.2024.0429000}

\title{Complex Permittivity Characterization of Low-Loss Dielectric Slabs at Sub-THz}

\author{\uppercase{Bing Xue}\authorrefmark{1}, \uppercase{Katsuyuki Haneda}\authorrefmark{1}, \IEEEmembership{Member,~IEEE}, \uppercase{Clemens Icheln}\authorrefmark{1}, \uppercase{Juha Ala-Laurinaho}\authorrefmark{1}, and \uppercase{Juha Tuomela}\authorrefmark{1}}

\address[1]{Department of Electronics and Nanoengineering, Aalto University-School of Electrical Engineering, Espoo FI-00076, Finland}
\tfootnote{The results presented in this paper have been supported by the Academy of Finland -- NSF joint call pilot ``Artificial intelligence and wireless communication technologies", decision \# 345178.}

\markboth
{Bing Xue \headeretal: Complex Permittivity Characterization of Low-Loss Dielectric Slabs at Sub-THz}
{Bing Xue \headeretal: Complex Permittivity Characterization of Low-Loss Dielectric Slabs at Sub-THz}

\corresp{Corresponding author: Bing Xue (e-mail: bing.xue@aalto.fi).}
\begin{abstract}
This paper presents a novel low-complexity method to characterize the permittivities of low-loss dielectric slabs at millimeter-wave and sub-terahertz bands. The method assumes plane-wave illumination of the material under test and utilizes reflected fields on two air-material interfaces, resolved through a sufficiently wide-band measurement. Unlike conventional methods, it does not need a reference measurement using a metal plate and a back metal plate. The applicability of the method is discussed. The misalignment of a reference plane and the effects of the gap between the back metal and the material under test are analyzed for published reflected-field based methods to show the advantages of the method. Finally, the proposed method is applied to estimate the permittivities of both plexiglass with 30 mm thickness and nylon with 21 mm thickness across 140-210 GHz and compared against the free-space transmission method, showing good agreement. The proposed method is suitable for permittivity characterization of low-loss materials when a wide-band plane-wave measurement is possible.
\end{abstract}


\titlepgskip=-21pt
\maketitle

\section{Introduction}
\IEEEPARstart{C}{omplex} permittivity stands as a crucial electrical characteristic of materials, with measurement technology tracing back to 100 years ago~\cite{von1954dielectrics}. Despite this historical foundation, characterizing permittivity remains a vibrant field, especially with the emergence of higher-precision measurement devices and the diversification of research areas, prompting a growing demand to comprehend complex permittivities across various frequency bands. In recent years, the sub-terahertz (sub-THz) bands have garnered attention for implementing imaging~\cite{wang2020efficient} and wireless communication~\cite{bicais2021}. Understanding the permittivities of materials has become pivotal for further research in these domains, prompting a shift in focus towards characterizing permittivity above $100$~GHz.

While various methodologies like optical resonator method~\cite{Degenford1966A}, time-domain spectroscopy systems~\cite{zhang2022quasi}, open-ended waveguides~\cite{christ2021reflection}, and waveguide-based transmission-reflection methods~\cite{shu2022characterisation} have been explored, free-space methods~\cite{Kazemipour2015Design, Yashchyshyn2018A, Zhu2021complex, zhang2023calibration, Taleb2023Transmission} have gained popularity because they offer accurate measurements by generating high-quality plane waves using quasi-optical components, i.e., parabolic mirrors and lenses. Free-space transmission methods in~\cite{Kazemipour2015Design, Zhu2021complex} have inherent complexity of the measurement setup and calibration, which is one important challenge of permittivity estimation at millimeter-wave (mmW) and sub-THz, since it requires at least two parabolic mirrors and two transceivers along with demanding extensive efforts for optical path calibration. The free-space reflection method at lower frequencies in~\cite{li2018compact, Hasar2022Complex, hasar2022broadband, zhang2023calibration} is not suited for sub-THz where precise measurement setups are necessary for, e.g., alignment of a reference metal plate, which necessitates expensive and precise fine displacement platforms and fixtures. This restricts the practical applications and increases the system's complexity;~\cite{Yashchyshyn2018A} proposed a novel method based on the free-space reflection coefficients without the knowledge of the material under test (MUT) thickness at sub-THz, unlike the majority of the free space methods that need to characterize the accurate thickness of MUT in~\cite{Kazemipour2015Design, Zhu2021complex,li2018compact, Hasar2022Complex, hasar2022broadband}. However, it still needs fine displacement platforms to align the reference metal plate, which influences the accuracy of permittivity estimates to some extent. Moreover, placing a metal plate behind the MUT can result in some unwanted gap in between, since the surfaces of MUT and metal plate cannot be considered perfectly flat relative to the short wavelength above 100 GHz. This further limits the practical applications of classical free-space reflection methods in~\cite{li2018compact, Hasar2022Complex, hasar2022broadband, zhang2023calibration} to sub-THz.~\cite{Hasar2022Calibration,Hasar2023dielectric} works on the time domain to obtain permittivities without a metal plate for calibrations. It used offset metal behind the MUT, avoiding the estimation errors from the unwanted gap, but it assumes no permittivities changes over frequencies. This limits its application to wide-band permittivity characterizations.~\cite{Guo2022Free} used two sample plates to avoid calibrating reflection coefficients. However, the distance between the two plates needs to be known accurately, and the transmission coefficient calibrations are still required. Furthermore, while many papers focus on characterizing the permittivity of thin materials at sub-THz, such as those discussed in~\cite{Kazemipour2015Design,Zhu2021complex}, there is continued interest in thicker materials, such as walls, glass, wood, concrete, and so on, as noted in~\cite{Hirata2021Measurement,Urahashi2021Complex,Aliouane2022Indoor}. The permittivities of these building materials are often utilized to investigate wireless channel characteristics at sub-THz.

This paper addresses the challenges of the complexity of the measurement setup and calibration. We propose a new and simple method by using wide-band free-space reflection coefficients of the MUT, avoiding using two parabolic mirrors and the second transceiver needed in the free-space transmission methods. The method does not need to use a metal plate at all, significantly reducing calibration complexity and estimate errors. Through full-wave simulations, we illustrate its advantages when comparing it to the ones with metal plates. We also show that our method characterizes the permittivity of low-loss dielectric slabs, with as good accuracy as the available free-space transmission method in~\cite{Zhu2021complex} by measurements.

\begin{figure}[htbp]
\centering
\includegraphics[width=0.800\linewidth]{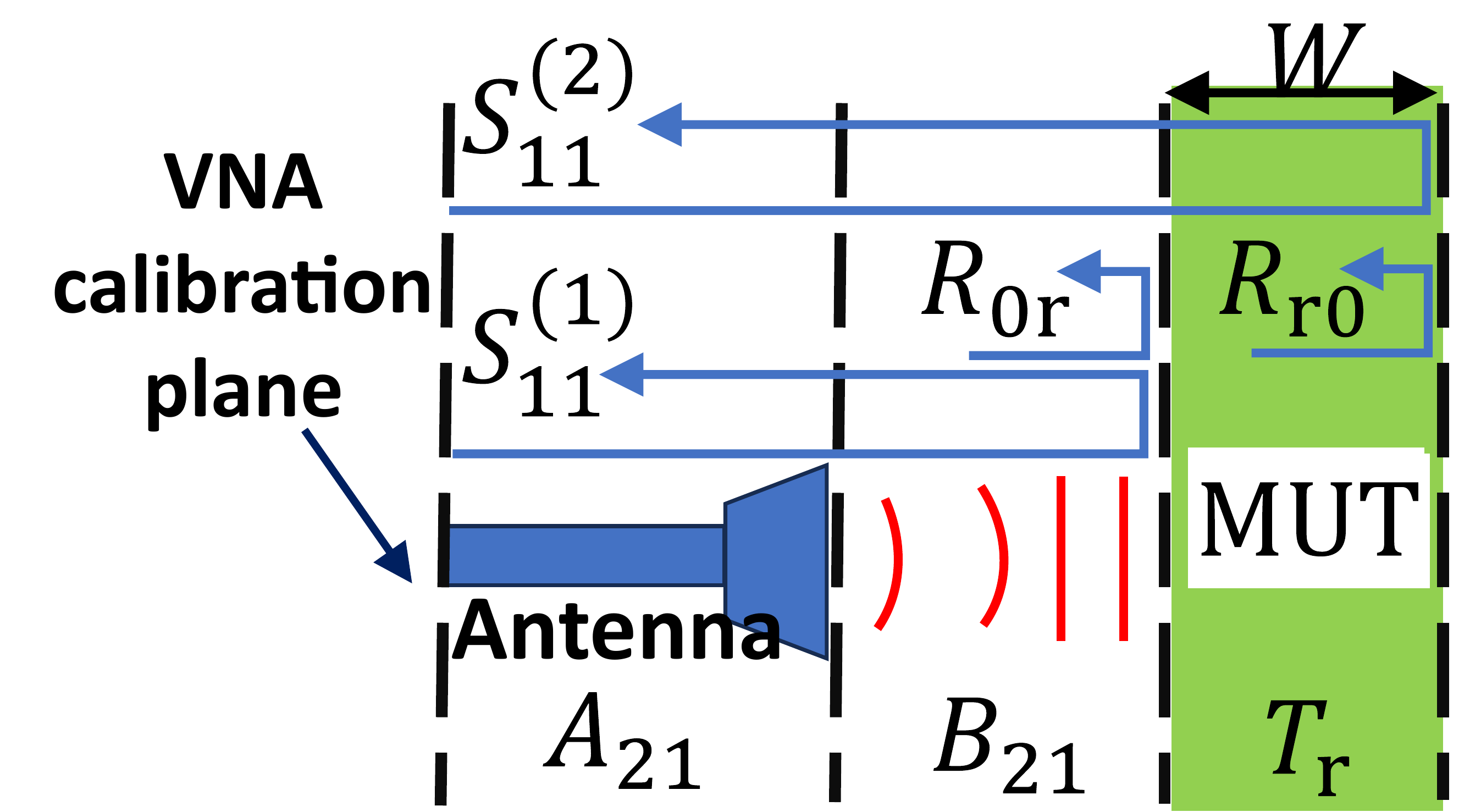}
\caption{Schematic of the measurement method.}
\label{fig:schematic}
\end{figure}
\section{Permittivity Characterization Method}
\label{sec:method}
\subsection{Algorithm}
\label{sec:algorithm1}
\begin{figure}[htbp]
\centering
\includegraphics[width=0.9\linewidth]{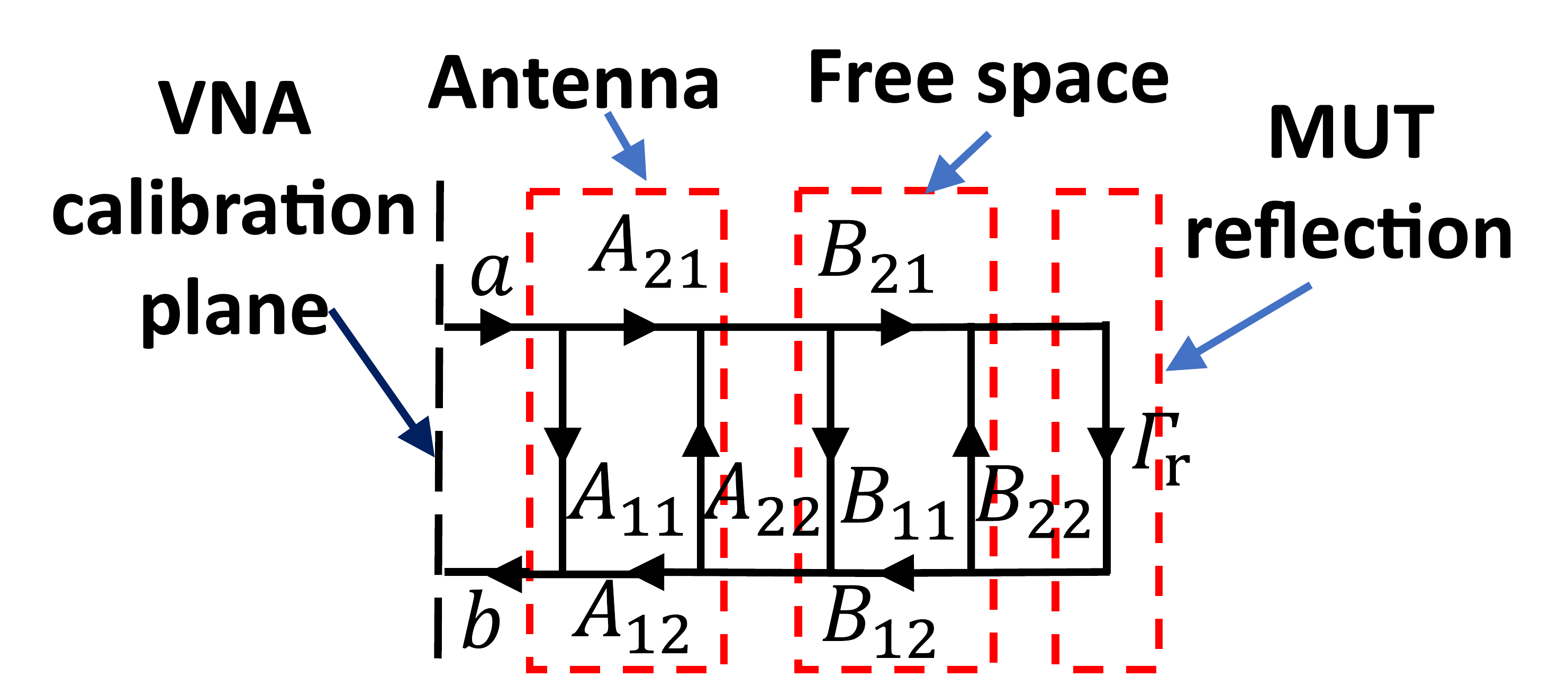}
\caption{Signal flow graph of the measurement system.}
\label{fig:signalflow}
\end{figure}

The schematic of the measurement method is shown in Fig.~\ref{fig:schematic}, where there is only one transceiver connecting the antenna, and the MUT is in the plane-wave zone. The relevant signal flow graph of the measurement system is depicted in Fig.~\ref{fig:signalflow}. 
The entire system operates under two key assumptions: 
1) the antenna functions as a two-port component and
2) no reflection occurs in the free space transmission (${B}_{\rm 11} = {B}_{\rm 22} = 0$). 
These assumptions allow the S-parameters measured at the vector network analyzer (VNA) to be expressed as follows:

\begin{equation}
\begin{aligned}
 {S}_{\rm 11} =  {A}_{\rm 11} + \frac{ {A}_{\rm 21} {B}_{\rm 21} \Gamma_{\rm r} {B}_{\rm 12} {A}_{\rm 12}}{1- {A}_{\rm 22} {B}_{\rm 21} \Gamma_{\rm r} {B}_{\rm 12}}, 
\label{eq:S11c}
\end{aligned}
\end{equation}
where $\Gamma_{\rm r}$ is the total reflection of the MUT. Matrices $[{\mathbf A}]^{2 \times 2}$ and $[{\mathbf B}]^{2 \times 2}$ represent the S-parameters of the antenna and the free space between a MUT and antenna. Due to the reciprocal nature of each component in the system, ${A}_{\rm 21} = {A}_{\rm 12}$ and ${B}_{\rm 21} = {B}_{\rm 12}$ is fulfilled. Here, ${A}_{\rm 22}$ signifies the structural scattering of the antenna and the antenna-transceiver mismatch. Typically, ${A}_{\rm 22}\Gamma_{\rm r} \ll 1$, and time-domain gating can be used to exclude the majority of multiple reflections between the antenna and the MUT, simplifying~\eqref{eq:S11c} as:
\begin{equation}
\begin{aligned}
 {S}_{\rm 11} =  {A}_{\rm 11} + {A}_{\rm 21}^2 {B}_{\rm 21}^2 \Gamma_{\rm r}.
\label{eq:S11s}
\end{aligned}
\end{equation}
When considering the reverberation of plane wave inside a MUT, it relates to the total reflection coefficient as 
\begin{equation}
\begin{aligned}
 \Gamma_{\rm r} = \Gamma_{\rm r,1} + \Gamma_{\rm r,2} + \cdots + \Gamma_{\rm r,\infty}, 
\label{eq:gamma}
\end{aligned}
\end{equation}
\begin{equation}
\begin{aligned}
 \Gamma_{\rm r,1} = R_{\rm 0r},\quad \Gamma_{{\rm r},n} = (1+ R_{\rm 0r})(1+ R_{\rm r0}) T_{\rm r}^{2n-2} R_{\rm r0}^{2n-3},
\label{eq:gammas}
\end{aligned}
\end{equation}
where $\Gamma_{{\rm r},n}$ denotes the $n$-th reflection within the MUT, and $n>1\cap n \in \mathbb{N}$; $R_{\rm 0r}$ and $R_{\rm r0}$ denote reflection coefficients at the two air-MUT interfaces as shown in Fig.~\ref{fig:schematic}, given by
\begin{equation}
R_{\rm 0r} = -R_{\rm r0} = \frac{1 -\sqrt{\varepsilon_{\rm r}}}{1+\sqrt{\varepsilon_{\rm r}}}.
\label{eq:Reflect}
\end{equation}
{where we assume air's relative permittivity is 1.} In~\eqref{eq:gammas}, the transmission coefficient $T_{\rm r}$ in MUT is written by
\begin{equation}
T_{\rm r} = \exp{\left(-{\rm i}k_0 W\sqrt{\varepsilon_{\rm r}}\right)},
\label{eq:Trans}
\end{equation}
{where we assume the MUT's relative permeability is 1, the same as the air.} $\varepsilon_{\rm r}$ represents the complex relative permittivity of MUT, typically expressed as $\varepsilon_{\rm r} = \varepsilon^{\prime} (1-{\rm i}\tan \delta)$, where $\varepsilon^{\prime}$ denotes the material's dielectric constant and $\tan \delta$ represents the dielectric loss tangent. Furthermore, in~\eqref{eq:Trans}, $W$ is its thickness, and $k_0$ is the wave number in free space. 

Combining~\eqref{eq:S11s} and~\eqref{eq:gamma}, S-parameters of plane waves reflected on the first and second air-MUT interfaces, defined in Fig.~\ref{fig:schematic}, can be described as
\begin{eqnarray}
\label{eq:S11_1}
 {S}_{\rm 11}^{(1)} & = & {A}_{\rm 21}^2 {B}_{\rm 21}^2 R_{\rm 0r}, \\
 {S}_{\rm 11}^{(2)} & = & -{A}_{\rm 21}^2 {B}_{\rm 21}^2 (1- R_{\rm 0r}^2) T_{\rm r}^2 R_{\rm 0r}.
\label{eq:S11_2}
\end{eqnarray}
By using~\eqref{eq:Reflect} and ~\eqref{eq:Trans}, and combining~\eqref{eq:S11_1} and~\eqref{eq:S11_2}, we can derive
\begin{equation}
\begin{aligned}
 \frac{{S}_{\rm 11}^{(2)}}{{S}_{11}^{(1)}} = -\frac{4\sqrt{\varepsilon_{\rm r}} }{\left(\sqrt{\varepsilon_{\rm r}} +1\right)^2}\exp{\left(-2{\rm i}k_0W\sqrt{\varepsilon_{\rm r}}\right)}.
\label{eq:glassSpa}
\end{aligned}
\end{equation}
where the factor ${A}_{\rm 21}^2 {B}_{\rm 21}^2$ disappears.
{When we assume $\tan \delta \ll 1$, the imaginary part of $-\frac{4\sqrt{\varepsilon_{\rm r}}}{\left(\sqrt{\varepsilon_{\rm r}} +1\right)^2}$ is negligible.} The total phase difference between ${S}_{\rm 11}^{(1)}$ and ${S}_{\rm 11}^{(2)}$ can be derived as:
\begin{equation}
\begin{aligned}
\Delta \phi &= \angle{\left(\frac{{S}_{\rm 11}^{(2)}}{{S}_{11}^{(1)}}\right)} + 2n\pi \\& = \angle{\left(-\frac{4\sqrt{\varepsilon_{\rm r}}}{\left(\sqrt{\varepsilon_{\rm r}} +1\right)^2}\exp{\left(-2{\rm i}k_0W\sqrt{\varepsilon_{\rm r}}\right)}\right)}+2n\pi \\& \approx \angle{\left(-\frac{4\sqrt{\varepsilon^{\prime}}}{\left(\sqrt{\varepsilon^{\prime}} +1\right)^2}\exp{\left(-2{\rm i}k_0W\sqrt{\varepsilon^{\prime}}\right)}\right)}+2n\pi\\& \approx \frac{4\sqrt{\varepsilon^{\prime}}}{\left(\sqrt{\varepsilon^{\prime}} +1\right)^2}\angle{\left(-\exp{\left(-2{\rm i}k_0W\sqrt{\varepsilon^{\prime}}\right)}\right)}+2n\pi
\label{eq:glassSpa1}
\end{aligned}
\end{equation}
where $\angle{\left(\cdot\right)}\in\left[0,2\pi\right)$ denotes an operator to determine the phase of a complex number, and $n \in \mathbb{N}$. Typically, $n$ can be estimated by 
\begin{eqnarray}
n = fix\left[\frac{2k_0W\sqrt{\hat{\varepsilon}^{\prime}}}{2\pi}\right]\label{eq:n}
\end{eqnarray}
where $\hat{\varepsilon}^{\prime}$ is the probable value of the dielectric constant of MUT. $fix[\cdot]$ denotes a function to obtain the integer part of a number.
Now we can derive $\varepsilon^{\prime}$ based on:
\begin{equation}
\begin{aligned}
 \varepsilon^{\prime} =\left(\frac{\Delta \phi}{2 k_0 W}\right)^2.
\label{eq:repart}
\end{aligned}
\end{equation}
Next, when $\tan \delta \ll 1$ in MUT, we can derive

\begin{equation}
\begin{aligned}
\sqrt{\varepsilon^{\prime} (1-{\rm i}\tan \delta)} \approx \sqrt{\varepsilon^{\prime}} \left(1-\frac{1}{2}{\rm i}\tan \delta  \right).
\label{eq:tand}
\end{aligned}
\end{equation}
Thus, $\left|{S}_{\rm 11}^{(2)} / {{S}_{11}^{(1)}}\right|$ can be represented as
\begin{equation}
\begin{aligned}
\left|\frac{{S}_{\rm 11}^{(2)}}{{S}_{11}^{(1)}}\right| \approx \frac{4\sqrt{\varepsilon^{\prime}}}{\left(\sqrt{\varepsilon^{\prime}} +1\right)^2}\exp{\left(-k_0W\sqrt{\varepsilon^{\prime}}\tan \delta \right)}.
\label{eq:abs}
\end{aligned}
\end{equation}
Consequently, $\tan \delta$ can be written as
\begin{equation}
\begin{aligned}
\tan \delta \approx -\frac{1}{k_0 W\sqrt{\varepsilon^{\prime}}}  \ln{\left(\frac{{\left(\sqrt{\varepsilon^{\prime}} +1\right)^2} }{4\sqrt{\varepsilon^{\prime}}}\left|\frac{{S}_{11}^{(2)} }{{S}_{11}^{(1)}}\right|\right)}.
\label{eq:impart}
\end{aligned}
\end{equation}

\begin{figure}[htbp]
    \centering
	\includegraphics[width=0.75\linewidth]{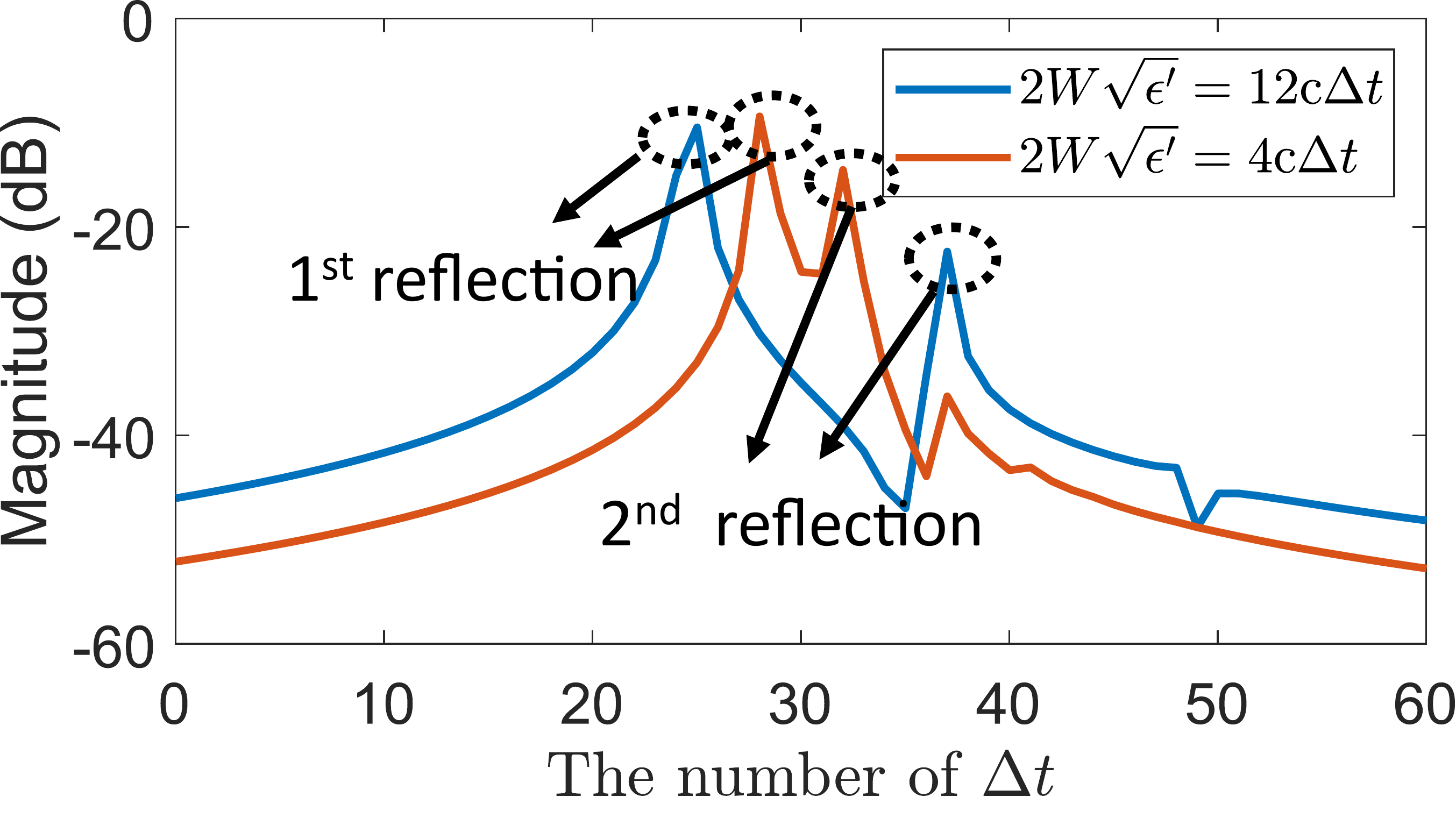}
	\caption{Impulse responses of reflection measurements for a MUT. Propagation delay of a reflection on the second air-MUT interface is longer than that of the first interface by $2W\sqrt{\varepsilon^{\prime}}= 4{\rm c} \Delta t$ and $2W\sqrt{\varepsilon^{\prime}}= 12{\rm c} \Delta t$, where $\rm c$ is the velocity of light.}
	\label{fig:timeMUT} 
\end{figure}

Estimation of ${S}_{11}^{(1)}$ and ${S}_{11}^{(2)}$ from the measured ${S}_{11}$, denoted as $\hat{S}_{11}$ hereinafter, relies on time-gating of impulse responses derived by the inverse discrete Fourier transform (IDFT) $\mathcal{F}^{-1}$ of $\hat{S}_{11}$~\cite{olkkonen2013Complex}. The time resolution is $\Delta t = 1/ B_0$ where $B_0$ is the measurement bandwidth set at the VNA. Resolving reflected plane waves on the first and second air-MUT interface in the time domain is challenging unless their time difference is much greater than $\Delta t$, i.e., the MUT thickness should be large enough to provide sufficient time delay. 
For instance, Fig.~\ref{fig:timeMUT} shows impulse responses derived from $\hat{S}_{11}$ where the propagation delay of a plane wave through the MUT is $4 \Delta t$ and $12 \Delta t$. \textbf{Algorithm 1} summarizes steps to estimate the permittivity of MUT from its measured $\hat{S}_{11}$ by VNA. Since the two reflected waves 
are closely spaced to each other in the time domain as exemplified in Fig.~\ref{fig:timeMUT}, the algorithm estimates them iteratively by search-and-subtraction for improved accuracy.
\begin{algorithm} [htbp]
\caption{Iterative Permittivity Estimation}
    \begin{algorithmic}[1]
    \STATE Obtain an initial estimate of a reflection on the first air-MUT interface: $\Tilde{S}_{\rm 11}^{(1,n)}(f) = \mathcal{F}\left\{K^{(1)}(t)\mathcal{F}^{-1}\left[\hat{S}_{\rm 11}(f)\right]\right\}$, where $\mathcal{F}\left[\cdot\right]$ is the DFT operator, $K^{(1)}(t)$ is a time-gating to extract the first reflection, and $n=0$ is an iteration index; $n \leftarrow n+1$.
    \STATE Subtract the reflection on the first interface from the measured S-parameter: $\Tilde{S}^{\prime}_{\rm 11}(f) = \hat{S}_{\rm 11}(f)-\Tilde{S}_{\rm 11}^{(1,n)}(f)$.
    \STATE Estimate a reflection on the second air-MUT interface using the first-reflection-cleaned S-parameter: $\Tilde{S}_{\rm 11}^{(2,n)}(f) = \mathcal{F}\left\{K^{(2)}(t)\mathcal{F}^{-1}\left[\Tilde{S}^{\prime}_{\rm 11}(f)\right]\right\}$, where $K^{(2)}(t)$ is the window to extract the second reflection.
    \STATE Subtract the second reflection from the measured S-parameter: $\Tilde{S}^{\prime\prime}_{\rm 11}(f) = \hat{S}_{\rm 11}(f)-\Tilde{S}_{\rm 11}^{(2,n)}(f)$.
    \STATE Obtain a refined estimate of the reflection on the first air-MUT interface using the second-reflection-cleaned S-parameter: $\Tilde{S}_{\rm 11}^{(1,n)}(f) = \mathcal{F}\left\{K^{(1)}(t)\mathcal{F}^{-1}\left[\Tilde{S}^{\prime\prime}_{\rm 11}(f)\right]\right\}$.
    \STATE Check if the refined estimate $\Tilde{S}_{\rm 11}^{(1,n)}(f)$ is identical to $\Tilde{S}_{\rm 11}^{(1,n-1)}(f)$. If yes, $\Tilde{S}_{\rm 11}^{(1)} = \Tilde{S}_{\rm 11}^{(1,n)}$ and $\Tilde{S}_{\rm 11}^{(2)} = \Tilde{S}_{\rm 11}^{(2,n)}$ and go to step 7; if no, $n \leftarrow n+1$ and go to step 2.
    \STATE Apply $\Tilde{S}_{\rm 11}^{(1)}$ and $\Tilde{S}_{\rm 11}^{(2)}$ to estimate complexity permittivity according to~\eqref{eq:glassSpa1},~\eqref{eq:repart}, and~\eqref{eq:impart}.
    \label{alg:1}
    \end{algorithmic}
\end{algorithm}
\subsection{Qualitative Analysis of the Method's Applicability}
\label{sec:Sensitivity}
\begin{figure}[htbp]
\centering
\subfigure[]{
\includegraphics[width=0.75\linewidth]{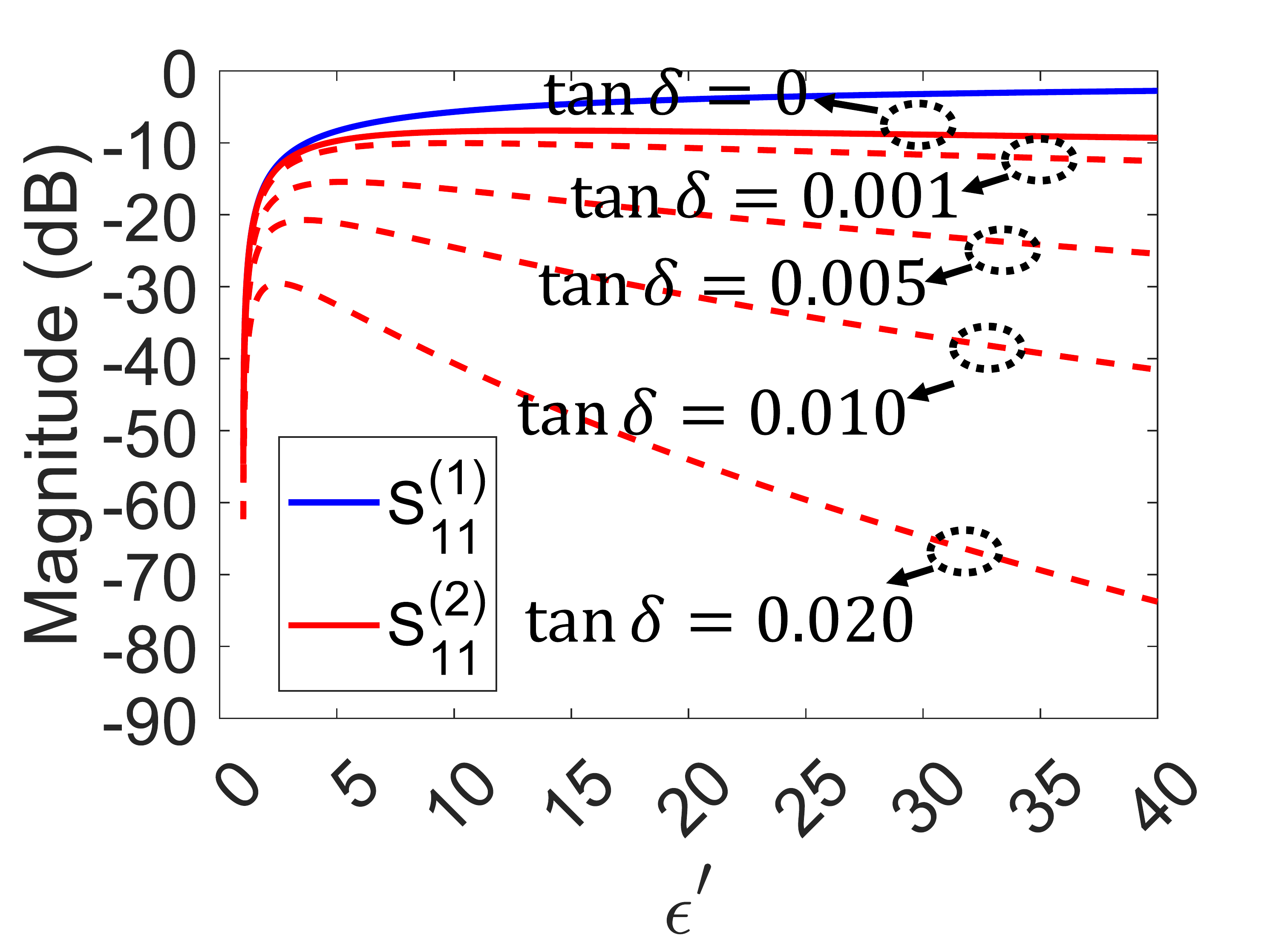}\label{fig:epchange}}
\subfigure[]{
\includegraphics[width=0.75\linewidth]{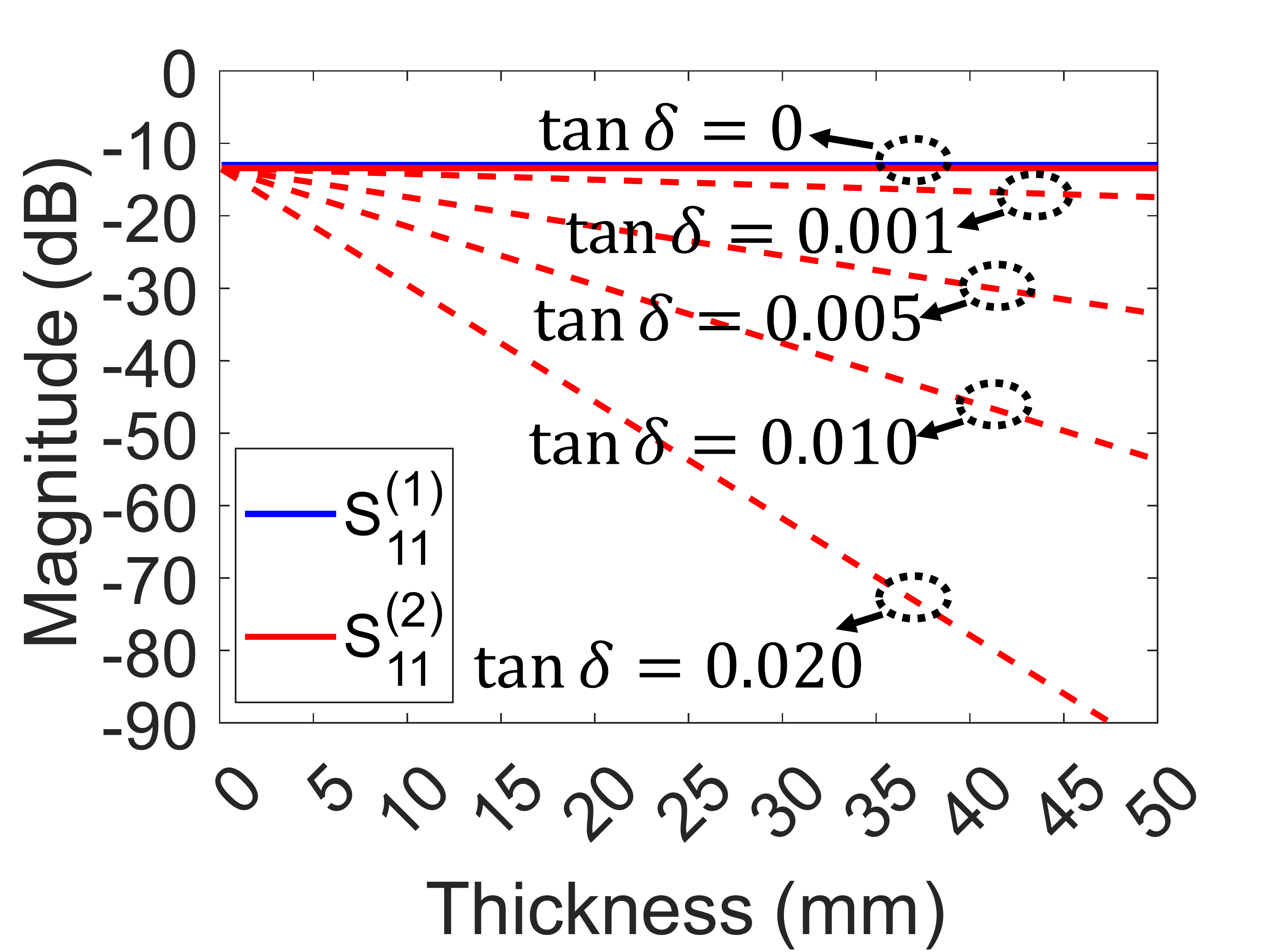}\label{fig:wchange}} 
\caption{(a) Magnitudes of ${S}_{\rm 11}^{(1)}$ and ${S}_{\rm 11}^{(2)}$ with respect to $\varepsilon^{\prime}$; $W = 10$~mm; $f=175$~GHz. (b) Magnitudes of ${S}_{\rm 11}^{(1)}$ and ${S}_{\rm 11}^{(2)}$ with respect to $W$; $\varepsilon^{\prime} =2.5$; $f=175$~GHz.}
\label{fig:epwchange}
\end{figure}

Observing~\eqref{eq:S11_1} and~\eqref{eq:S11_2}, we notice the inverse proportionality between ${S}_{\rm 11}^{(1)}$ and ${S}_{\rm 11}^{(2)}$. Assuming ${A}_{\rm 21}^2 {B}_{\rm 21}^2 \approx 1$ and $R_{\rm 0r} \approx (1 -\sqrt{\varepsilon^{\prime}})/(1+\sqrt{\varepsilon^{\prime}})$ due to $\tan \delta \ll 1$, we calculated $|{S}_{\rm 11}^{(1)}|$ and $|{S}_{\rm 11}^{(2)}|$ for varying $\varepsilon^{\prime}$ at $W = 10$~mm and $f=175$~GHz, as shown in Fig.~\ref{fig:epchange}. The figure reveals an increasing difference between $|{S}_{\rm 11}^{(1)}|$ and $|{S}_{\rm 11}^{(2)}|$ with higher values of $\varepsilon^{\prime}$. Additionally, higher $\tan \delta$ further amplifies this difference. Furthermore, we computed the magnitudes of ${S}_{\rm 11}^{(1)}$ and ${S}_{\rm 11}^{(2)}$ for varying $W$ at $\varepsilon^{\prime} =2.5$ and $f=175$~GHz, depicted in Fig.~\ref{fig:wchange}. A similar trend as Fig.~\ref{fig:epchange} can be found. When these reflection coefficients are close to the noise floor of a VNA, e.g., $-100$~dB, the permittivity estimates will lose accuracy. Moreover, Fig.~\ref{fig:timeMUT} also explains that the effective electric length ($W\sqrt{\varepsilon_{\rm r}}$) of the MUT should be much larger than the time resolution of an impulse response to separate ${S}_{\rm 11}^{(1)}$ and ${S}_{\rm 11}^{(2)}$ properly. Therefore, the proposed method is mainly suitable for permittivity characterizations of low-loss thick slabs.
\subsection{Numerical Simulations}
Full-wave simulations of plane wave incidence on a finite-thickness MUT are performed to evaluate the efficacy of the proposed method relying on reflection measurements. The simulations assume a noise-free scenario. In the implementation of our method, \textbf{Algorithm 1} uses the Kaiser-Bessel window with $\beta = 6$ for time-gating, showing a low side lobe level in the frequency domain. We also include the transmission method in~\cite{Zhu2021complex} as a benchmark for our method validation. Both methods rely on resolving the signals in the time domain. 

Moreover, we also study permittivity estimation errors due to the misalignment of the reference plane and metal-MUT gap for the classical reflection methods in~\cite{Yashchyshyn2018A} and in~\cite{hasar2022broadband}, respectively. 
\subsubsection{Algorithm Validation}
\begin{figure}[htbp]
\centering
\subfigure[]{
\includegraphics[width=0.75\linewidth]{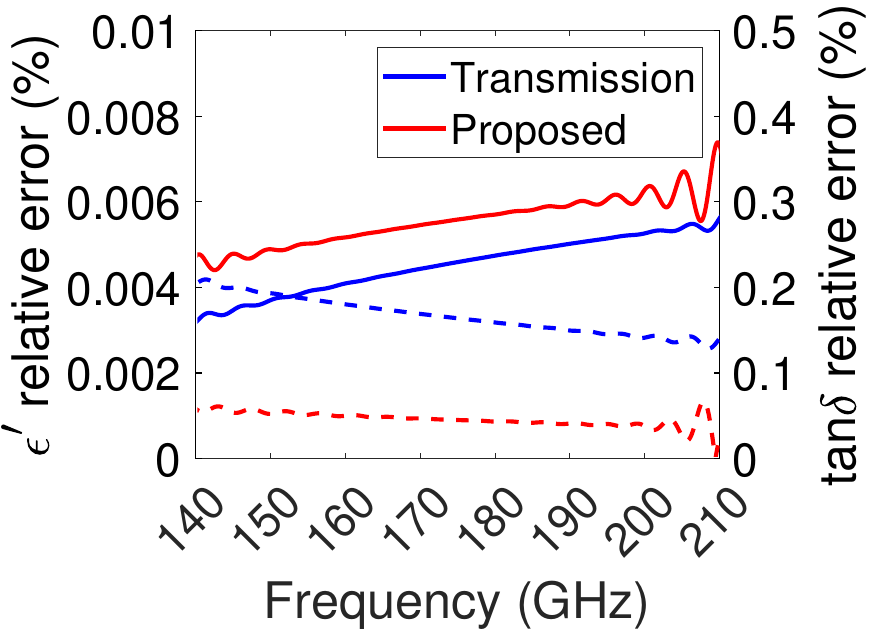}\label{fig:example_1}}
\subfigure[]{
\includegraphics[width=0.75\linewidth]{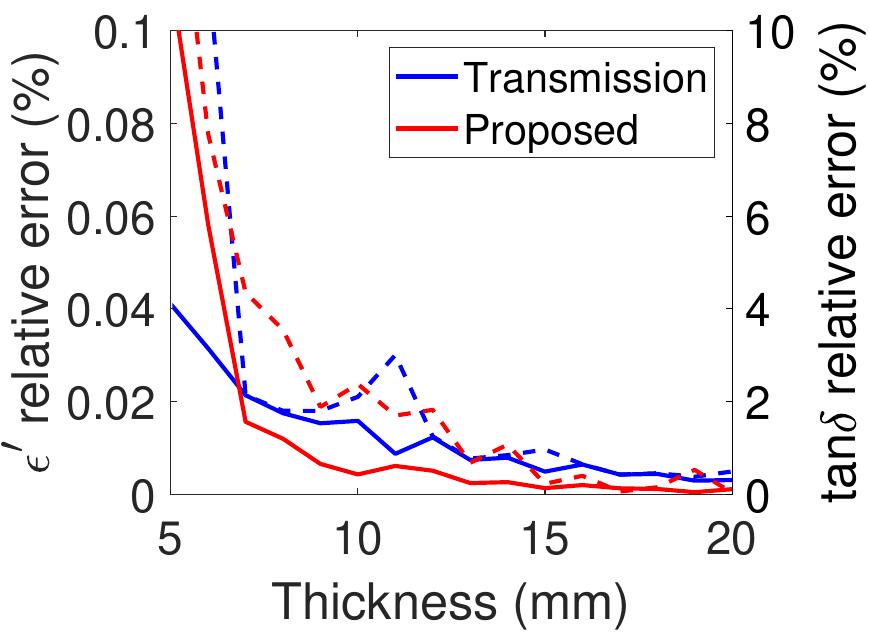}\label{fig:example_2}} 
\caption{Relative error of estimates based on simulated S-parameters of a plane wave on MUT (a) for validation (b) for different MUT thicknesses. The solid lines are for $\varepsilon^{\prime}$ relative error, and the dashed lines are for $\tan \delta$ relative error.}
\label{fig:examples}
\end{figure}

$\varepsilon_{\rm r} = 5\times(1-0.02\rm i)$ and $W=30$~mm. The S-parameters are obtained between $130$ and $220$~GHz, yielding $\Delta t = 11.1$~ps. The window width for time gating is $40 \Delta t$ for both the proposed method and the transmission method in~\cite{Zhu2021complex}. The permittivity estimates at the upper- and lower-most $10$~GHz bands suffer from insufficient accuracy due to the limited window width to resolve the two reflections properly, making the effective frequency range between $140$ and $210$~GHz. Then, we calculate the relative error of the two methods along the frequencies. The relative error ${err} = \frac{\left|x-\bar{x}\right|}{x}$ is evaluated where $\bar x$ represents the estimate of $\varepsilon^{\prime}$ or $\tan \delta$, while $x$ denotes their true values. Fig.~\ref{fig:example_1} shows that the proposed method estimates $\varepsilon^{\prime}$ and $\tan \delta$ accurately as the transmission method~\cite{Zhu2021complex} does. 
\subsubsection{MUT Thickness Influence on Estimate Accuracy}
Let us set $\varepsilon_{\rm r} = 3\times (1-0.01\rm i)$, and change $W$ from 5~mm to 20~mm. The relative errors of estimates for each $W$ value are obtained by averaging them over the range $140$ and $210$~GHz. Fig.~\ref{fig:example_2} depicts $err$, showing that our method has a similar accuracy as the transmission method in~\cite {Zhu2021complex}. However, the estimation errors of our method and the transmission method both increase when the MUT is thin, especially for the loss tangent. This means our method is major suitable for thick materials. Or wider frequency bandwidth can help to characterize thin materials' permittivities by increasing temporal resolution. 
\subsubsection{Imperfect Alignment of the Reference plane}
\begin{figure}[htbp]
\centering
\subfigure[]{
\includegraphics[width=0.75\linewidth]{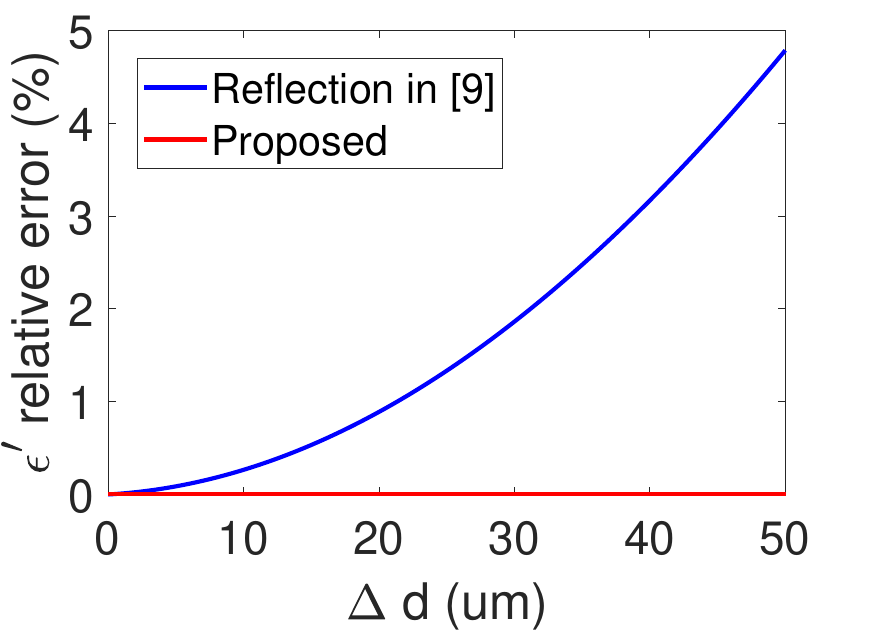}
}
\subfigure[]{
\includegraphics[width=0.75\linewidth]{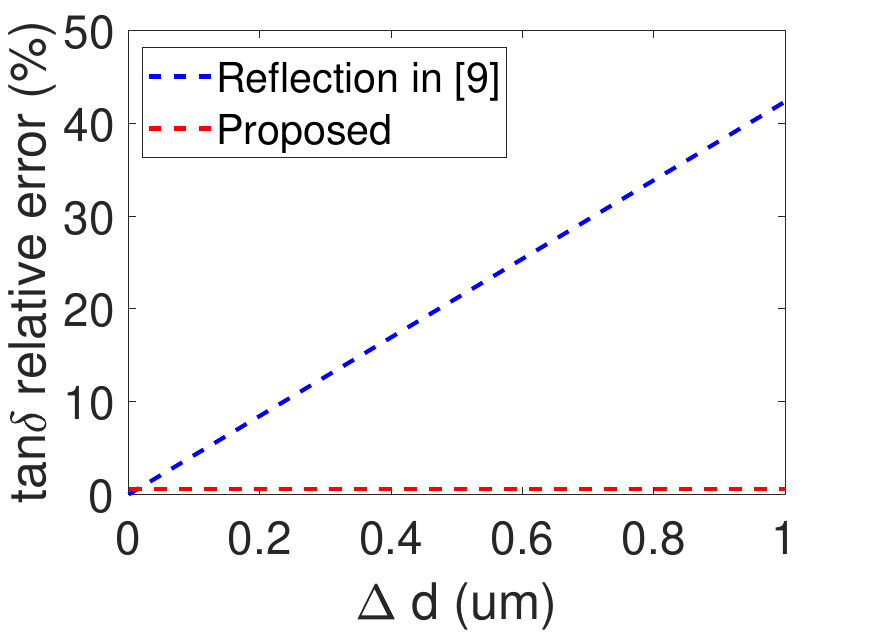}
} 
\caption{Relative error of estimates based on simulated S-parameters of a plane wave on MUT with $\Delta d$ changes (a) for $\varepsilon^{\prime}$ and (b) for $\tan \delta$.}
\label{fig:example_3}
\end{figure}

For the reflection methods that need reference-plane calibrations, the metal plate is often used for this purpose, e.g.,~\cite{Yashchyshyn2018A}. $\Delta d$ indicates the distance between the reference plane and MUT's near surface; $\Delta d =0$ means perfect alignment. Let us set $\varepsilon_{\rm r} = 3\times (1-0.01\rm i)$ and $W=30$~mm. The relative errors of estimates for each $\Delta d$ value are obtained by averaging them over the range $140$ to $210$~GHz. When changing $\Delta d$, we can see from Fig.~\ref{fig:example_3} that $\tan \delta$ is more sensitive to the alignment of the reference plane than $\varepsilon^{\prime}$. Even $\Delta d=1~\rm\mu $m, $\tan \delta$ relative error is over $40$\%. Our method can avoid this issue, reducing the usage of expensive precise displacement platforms.

\subsubsection{Metal-MUT Gap Influences}
\begin{figure}[htbp]
\centering
\subfigure[]{
\includegraphics[width=0.75\linewidth]{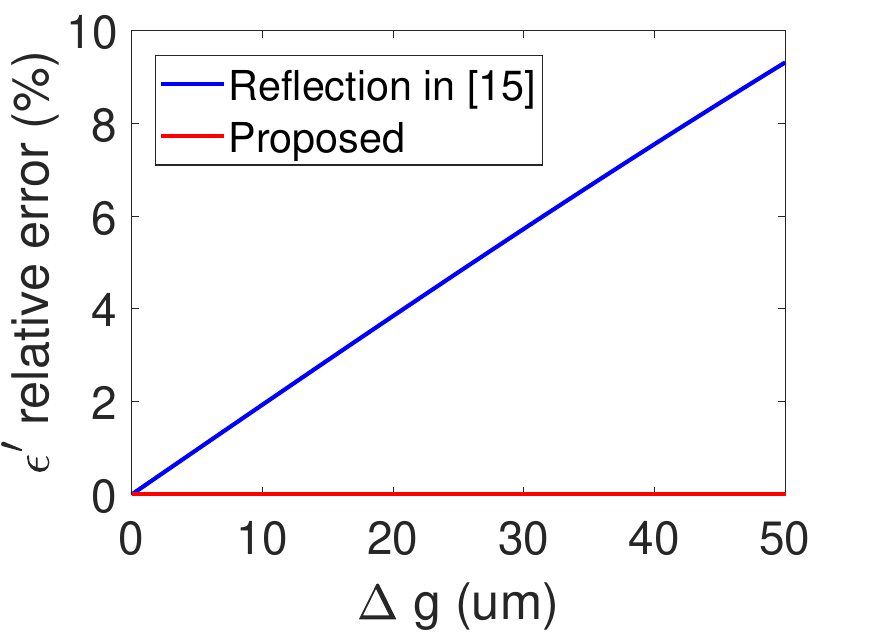}
}
\subfigure[]{
\includegraphics[width=0.75\linewidth]{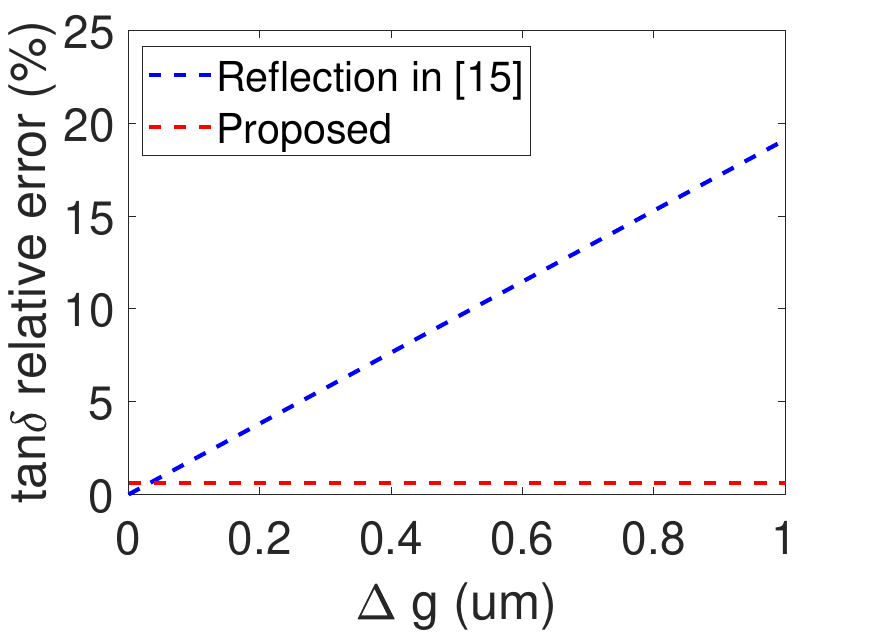}
} 
\caption{Relative error of estimates based on simulated S-parameters of a plane wave on MUT with $\Delta g$ changes (a) for $\varepsilon^{\prime}$ and (b) for $\tan \delta$.}
\label{fig:example_4}
\end{figure}

Another significant error of permittivity estimation comes from the gap between MUT and the back metal when relying on the reflection coefficients on the interface~\cite{hasar2022broadband}; $\Delta g$ indicates this gap. Let us set $\varepsilon_{\rm r} = 3\times (1-0.01\rm i)$ and $W=30$~mm. The relative errors of estimates for each $\Delta g$ value are obtained by averaging them over the range $140$ to $210$~GHz. We can see from Fig.~\ref{fig:example_4} even $\Delta g=1~\rm\mu $m, $\tan \delta$ relative error is over $19$\%. Our method can avoid this issue, reducing the estimate errors because of the metal-MUT gap coming from the imperfect surfaces of the metal plate and MUT.

\section{Experiment}
\subsection{Quasi-optical System}
\begin{figure}[htbp]
    \centering
	\includegraphics[width=0.7\linewidth]{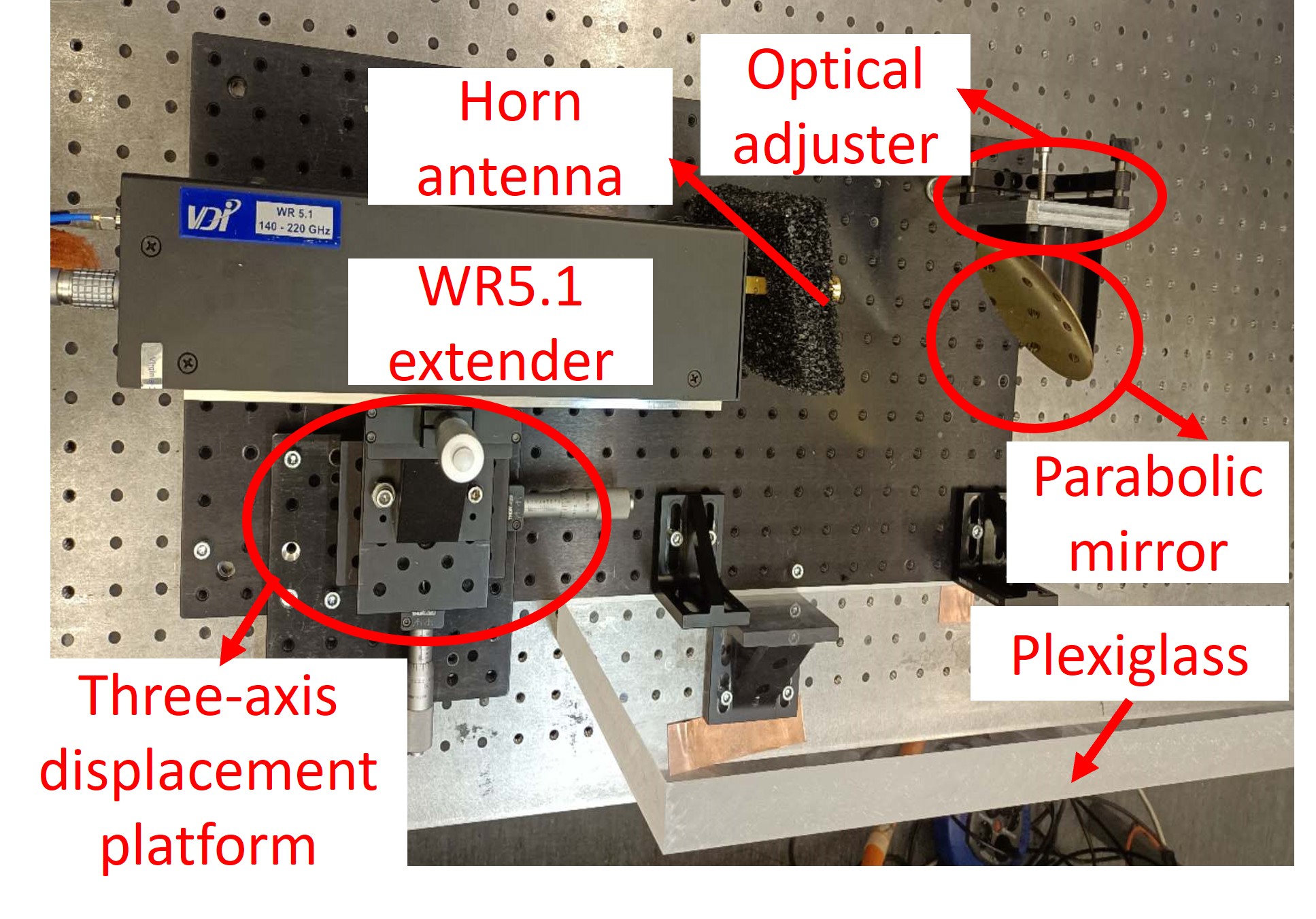}
	\caption{The one-mirror setup for plexiglass permittivity measurement.}
	\label{fig:glasssetup}
\end{figure}

In the quasi-optical system illustrated in Fig.~\ref{fig:glasssetup}, we employed a corrugated conical horn antenna with a gain greater than $20$~dBi as the feed to achieve high Gaussicity of the beam. An off-axis parabolic mirror with a focal length of $152.4~\rm m$m and a beam reflection angle of $90^\circ$ was utilized to collimate the beam and create a planar wave. The wave planarity was verified with a near-field scanner, and a three-axis displacement platform and an optical adjuster were used to tune the feed and mirror positions. Our one-mirror system has smaller free-space losses than that of the two-mirror system built in~\cite{Zhu2021complex} by $3.5$~dB at $140~$~GHz and by $0.3$~dB at $220~$~GHz. In our setup, the frequency range was configured from $130~$~GHz to $220~$~GHz with the intermediate frequency $1$~kHz. To mitigate aliasing due to the multiple reflections between the antenna and MUT and the environment reflections in the impulse responses, $1601$ sampling points were used.

\subsection{Results}
\begin{figure}[htbp]
\centering
	\includegraphics[width=0.75\linewidth]{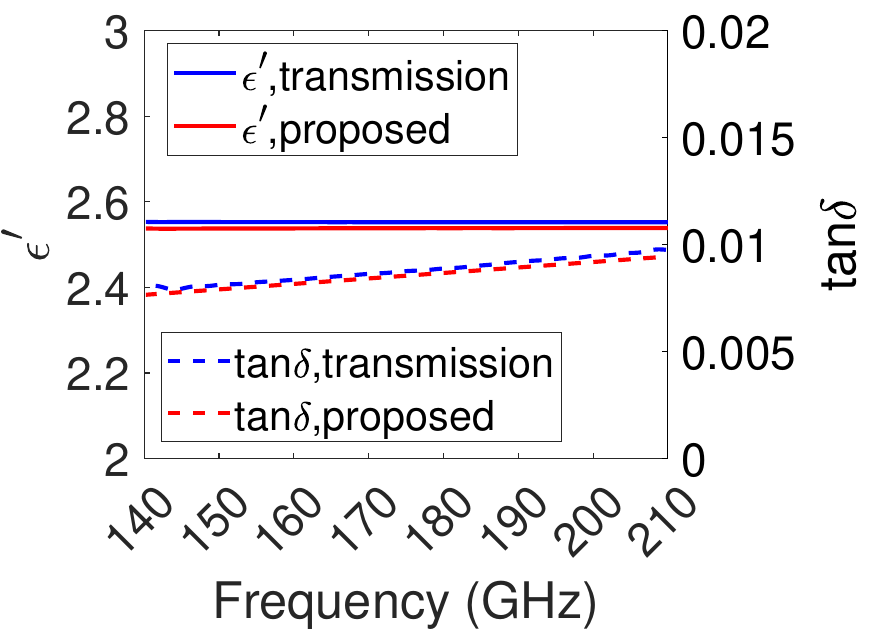}
\caption{{Mean permittivity of the plexiglass plate across the frequency range of 140-210 GHz.}}
\label{fig:glassDK}
\end{figure}

\begin{figure}[htbp]
\centering
\includegraphics[width=0.75\linewidth]{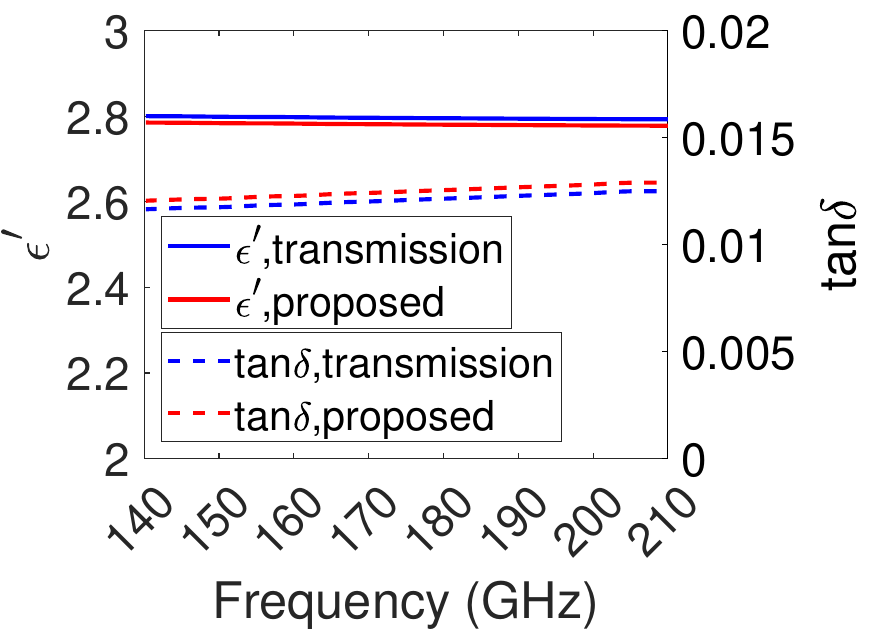}
\caption{{Mean permittivity of the nylon plate across the frequency range of 140-210 GHz.}}
\label{fig:PTFEDK}
\end{figure}

A $30$-mm thick plexiglass with flat and smooth surfaces was positioned within the plane wave zone of the Gaussian beam. {A caliper with $0.01$~mm resolution was used to measure the thickness of the illuminated area 10 times, the average value is $W =29.65$~mm.} After measuring S-parameters, our proposed method was applied to derive the permittivity using a window width of $40 \Delta t$. We also conducted permittivity measurements of the plexiglass using the transmission method~\cite{Zhu2021complex}. Each method underwent $10$ S-parameter measurements, and the calculated mean estimates are shown in Fig.~\ref{fig:glassDK}. The maximum difference {between these two methods} in $\varepsilon^{\prime}$ was $1.1\times 10^{-2}$, whereas that of $\tan \delta$ was $7.1\times 10^{-4}$ across the frequency range of 140-210 GHz. 
A $21$~mm thick nylon plate with flat and smooth surfaces was measured using the same setup as plexiglass. {A caliper with 0.01 mm resolution was used to measure the thickness of the illuminated area 10 times. The average value of the estimate is $W = 29.65$ mm.} We can find that the maximum difference between these two methods in $\varepsilon^{\prime}$ was $1.3\times 10^{-2}$, whereas that of $\tan \delta$ was $4.5\times 10^{-4}$ across the frequency range of 140-210 GHz as shown in Fig.~\ref{fig:PTFEDK}. 

In Table~\ref{uncertainty2}, we compare our estimated permittivity of plexiglass with that from~\cite{Degenford1966A} only available at 143 GHz and from the transmission method in~\cite{Zhu2021complex}, showing agreement to each other. The averaged estimation difference between the proposed method and the value provided in~\cite{Degenford1966A} is around $2.3\%$ probably due to the accuracy of the thickness measurement of the plexiglass slab and permittivity difference between two different samples.
\begin{table}[htbp]
	\begin{center}
		\caption{Plexiglass Permittivity Comparisons} 
		\label{uncertainty2}
  \begin{tabular}{c|c|r}\hline\hline 
  \multirow{2}{*}{Proposed}&$\varepsilon^{\prime}$ &$2.54$ ($143$~GHz)\\\cline{2-3}
   &$\tan \delta$&$0.0077$ ($143$~GHz)\\\hline
    \multirow{2}{*}{Transmission}&$\varepsilon^{\prime}$ &$2.55$ ($143$~GHz)\\\cline{2-3}
   &$\tan \delta$&$0.0080$ ($143$~GHz)\\\hline
   \multirow{2}{*}{Reference~\cite{Degenford1966A}}&$\varepsilon^{\prime}$ &$2.60$ ($143$~GHz)\\\cline{2-3}
   &$\tan \delta$&$-$ \\\hline\hline
 	\end{tabular}
	\end{center}   
	\end{table}
 
In Table~\ref{uncertainty3}, we compare our estimated permittivity of Nylon with that from the transmission method and its reference value at 26-40 GHz. It can be observed that the proposed method yields a dielectric constant close to those obtained by the transmission method and the reference value in~\cite{Abbas2001Complex}. The difference in loss tangent between the proposed method and the reference value is attributed to differences in the samples and frequencies.
\begin{table}[htbp]
	\begin{center}
		\caption{Nylon Permittivity Comparisons} 
		\label{uncertainty3}
  \begin{tabular}{c|c|r|r}\hline\hline 
  \multirow{2}{*}{Proposed}&$\varepsilon^{\prime}$ &$2.79$ ($140$~GHz)&$2.78$ ($210$~GHz)\\\cline{2-4}
   &$\tan \delta$&$0.00116$ ($140$~GHz)&$0.0125$ ($210$~GHz)\\\hline
    \multirow{2}{*}{Transmission}&$\varepsilon^{\prime}$ &$2.80$ ($140$~GHz)&$2.79$ ($210$~GHz)\\\cline{2-4}
      &$\tan \delta$&$0.0121$ ($140$~GHz)&$0.0129$ ($210$~GHz)\\\hline
   \multirow{2}{*}{Reference~\cite{Abbas2001Complex}}&$\varepsilon^{\prime}$ &$3.00$ ($26-40$~GHz)\\\cline{2-4}
   &$\tan \delta$&$0.018$  ($26-40$~GHz) \\\hline\hline
 	\end{tabular}
	\end{center}   
	\end{table}
 
It is noted that, based on~\eqref{eq:repart}, the error of dielectric constant estimate $\Delta\varepsilon^{\prime}$ is related to that of the slab thickness measurement $\Delta W$ as
\begin{equation}
\begin{aligned}
 \Delta\varepsilon^{\prime} \approx \frac{\left(\Delta \phi\right)^2}{2 k_0^2 W^3}\Delta W.
\label{eq:errDK}
\end{aligned}
\end{equation}
Then, according to~\eqref{eq:impart}, $\Delta W$ results in an error of loss tangent $\Delta\tan \delta$, written as
\begin{equation}
\begin{aligned}
\Delta\tan \delta \approx \frac{1}{k_0 W^2\sqrt{\varepsilon^{\prime}}}  \ln{\left(\frac{{\left(\sqrt{\varepsilon^{\prime}} +1\right)^2}}{4\sqrt{\varepsilon^{\prime}}}\left|\frac{{S}_{11}^{(2)} }{{S}_{11}^{(1)}}\right|\right)}\Delta W.
\label{eq:errtan}
\end{aligned}
\end{equation}
Both~\eqref{eq:errDK} and~\eqref{eq:errtan} indicate that thicker MUT can reduce the estimation error due to $\Delta W$. 

\subsection{Uncertainty}
{To estimate the uncertainty, we consider the major sources of error in, e.g., sample placement, sample thickness measurement, and the measured S-parameters.}

\subsubsection{Placement}
The permittivity of the samples was estimated by placing the sample in the holder 10 times. The standard uncertainty in the obtained permittivity values was then calculated using (8), (9) of~\cite{Wang2020Characterization}. The results for the plexiglass sample are listed in Table~\ref{uncertainty1}.

\begin{table}[htbp]
	\begin{center}
		\caption{{Permittivity Uncertainty Due to Different Error Sources for Plexiglass}} 
		\label{uncertainty1}
  \begin{tabular}{c|c|c|r|r|r|r}\hline\hline 
 \multicolumn{3}{c|}{Frequency}&140GHz&160GHz&180GHz&200GHz\\\hline\hline
  \multirow{8}{*}{Tra}&\multirow{2}{*}{P}&$\varepsilon^{\prime}$&0.011&0.012&0.011&0.011\\\cline{3-7}
  &&$\tan \delta$&0.00052&0.00054&0.00048&0.00053 \\\cline{2-7}
&\multirow{2}{*}{T}&$\varepsilon^{\prime}$&0.025&0.026&0.024&0.022 \\\cline{3-7}
  &&$\tan \delta$&0.00014&0.00021&0.00017&0.00024\\\cline{2-7}
&\multirow{2}{*}{S}&$\varepsilon^{\prime}$&0.00061&0.00056&0.00058&0.00064\\\cline{3-7}
  &&$\tan \delta$&0.000061&0.000055&0.000055&0.000059 \\\cline{2-7}
 &\multirow{2}{*}{Tot}&$\varepsilon^{\prime}$&0.027&0.029&0.026&0.025\\\cline{3-7}
  &&$\tan \delta$&0.00053&0.00058&0.00051&0.00058 \\\hline
  \multirow{8}{*}{Pro}&\multirow{2}{*}{P}&$\varepsilon^{\prime}$&0.014&0.012&0.013&0.015\\\cline{3-7}
  &&$\tan \delta$&0.00057&0.00058&0.00061&0.00057 \\\cline{2-7}
&\multirow{2}{*}{T}&$\varepsilon^{\prime}$&0.041&0.038&0.042&0.044 \\\cline{3-7}
  &&$\tan \delta$&0.00014&0.00015&0.00014&0.00012\\\cline{2-7}
&\multirow{2}{*}{S}&$\varepsilon^{\prime}$&0.00075&0.00078&0.00078&0.00074\\\cline{3-7}
  &&$\tan \delta$&0.000051&0.000054&0.000055&0.000054 \\\cline{2-7}
 &\multirow{2}{*}{Tot}&$\varepsilon^{\prime}$&0.044&0.041&0.045&0.047\\\cline{3-7}
  &&$\tan \delta$&0.00059&0.00060&0.00063&0.00058 \\\hline\hline
 	\end{tabular}
	\end{center}
 Pro: proposed method; Tra: transmission method in~\cite{Zhu2021complex}; P: placement; T: thickness; S: S-parameter; Tot: total uncertainty.
	\end{table}

\subsubsection{Measurements of Thickness}
The sample thickness was measured using a caliper with a resolution of $0.01$~mm. The standard uncertainty due to the caliper's measurement error can be approximated as $10/\sqrt{3}\approx 6$~$\mu$m~\cite{Wang2020Characterization}. Additionally, the thickness variation within the illuminated area of the sample was estimated by measuring the thickness 10 times. According to Section 4.2 of ~\cite{GUM2008}, the uncertainty in thickness due to variation within the sample for a $30$-mm plexiglass sample was estimated to be $0.22$~mm. Therefore, the total uncertainty in permittivity values due to thickness measurements for the plexiglass sample was calculated using (10) of Section 5.1.2 in ~\cite{GUM2008}, and the results are listed in Table~\ref{uncertainty1}.

\subsubsection{S-Parameter Measurements}
S-parameter measurements are generally influenced by various errors, including VNA system measurement errors, calibration, environmental influences and so on~\cite{Wang2020Characterization}. The transmission method~\cite{Zhu2021complex} characterizes permittivity primarily based on the differences between $S_{21}$ with and without samples, whereas the proposed method relies on the differences between the first and second surface reflections of the samples. Both methods are self-calibrating, as they eliminate propagation coefficients ${A}_{\rm 21}^2 {B}_{\rm 21}^2$ in the final formulas, such as~\eqref{eq:glassSpa} and (9) in~\cite{Zhu2021complex}. Thus, we assume the main error sources for both methods arise from systematic errors of VNA measurements and environmental influences. The uncertainty in the S-parameters was calculated using the \textit{Keysight Uncertainty Calculator}~\cite{Keysight2023}. Consequently, the total uncertainty in permittivity values due to S-parameter measurements for the plexiglass sample was calculated using (10) of Section 5.1.2 in~\cite{Wang2020Characterization}, with the results shown in Table~\ref{uncertainty1}.

\subsubsection{Total Uncertainty}
The total uncertainty for the plexiglass sample was combined using (10) in~\cite{GUM2008}. As presented in Table~\ref{uncertainty1}, the proposed method exhibits higher uncertainty in both thickness and S-parameter measurements than the transmission method. Regarding the total uncertainty, the proposed method demonstrates similar levels of uncertainty for loss tangents but higher uncertainty for dielectric constants than the transmission method.
 
\section{Conclusions}
\label{sec:conclusion}
The paper introduces a permittivity characterization method relying on accurate measurements of plane wave reflections on two air-MUT interfaces. The method can accurately characterize a thick low-loss slab having flat and smooth surfaces with high accuracy. However, the accuracy is reduced for thin materials. Compared with the transmission method in 140-210 GHz, our method shows similar estimation values and uncertainties for characterizing plexiglass, despite our method requiring lower system complexity and reduced measurement efforts. Compared with the classical free-space reflection methods, our method avoids a back metal plate and an alignment of a reference metal plate, which can easily cause huge estimate errors at sub-THz. Hence, our method is a strong candidate for permittivity characterizations at mmW and sub-THz, when MUT has suitable properties and wideband plane-wave illumination of MUT is feasible.

\bibliographystyle{IEEEtran}
\footnotesize
\bibliography{refrealhand}

\begin{thebibliography}{10}
\providecommand{\url}[1]{#1}
\csname url@samestyle\endcsname
\providecommand{\newblock}{\relax}
\providecommand{\bibinfo}[2]{#2}
\providecommand{\BIBentrySTDinterwordspacing}{\spaceskip=0pt\relax}
\providecommand{\BIBentryALTinterwordstretchfactor}{4}
\providecommand{\BIBentryALTinterwordspacing}{\spaceskip=\fontdimen2\font plus
\BIBentryALTinterwordstretchfactor\fontdimen3\font minus \fontdimen4\font\relax}
\providecommand{\BIBforeignlanguage}[2]{{%
\expandafter\ifx\csname l@#1\endcsname\relax
\typeout{** WARNING: IEEEtran.bst: No hyphenation pattern has been}%
\typeout{** loaded for the language `#1'. Using the pattern for}%
\typeout{** the default language instead.}%
\else
\language=\csname l@#1\endcsname
\fi
#2}}
\providecommand{\BIBdecl}{\relax}
\BIBdecl

\bibitem{von1954dielectrics}
R.~Von~Hipple, ``Dielectrics and waves,'' \emph{New York: John Willey and Sons}, 1954.

\bibitem{wang2020efficient}
C.~Wang, Q.~Zhang, J.~Hu, C.~Li, S.~Shi, and G.~Fang, ``An efficient algorithm based on csa for thz stepped-frequency sar imaging,'' \emph{IEEE Geoscience and Remote Sensing Letters}, vol.~19, pp. 1--5, 2020.

\bibitem{bicais2021}
S.~Bica{\"\i}s, A.~Falempin, J.-B. Dor{\'e}, and V.~Savin, ``{Design and Analysis of MIMO Systems using Energy Detectors for Sub-THz Applications},'' \emph{IEEE Transactions on Wireless Communications}, vol.~21, no.~6, pp. 3678--3690, 2021.

\bibitem{Degenford1966A}
J.~Degenford and P.~Coleman, ``A quasi-optics perturbation technique for measuring dielectric constants,'' \emph{Proceedings of the IEEE}, vol.~54, no.~4, pp. 520--522, 1966.

\bibitem{zhang2022quasi}
Z.~Zhang, R.~Jia, J.~Xu, Q.~Niu, Y.~Yang, C.~Zhang, and Y.~Zhao, ``Quasi-optical measurement and complex refractive index extraction of flat plate materials using single time-domain transmission model in y-band,'' \emph{IEEE Transactions on Microwave Theory and Techniques}, vol.~70, no.~11, pp. 5224--5233, 2022.

\bibitem{christ2021reflection}
A.~Christ, A.~Aeschbacher, F.~Rouholahnejad, T.~Samaras, B.~Tarigan, and N.~Kuster, ``Reflection properties of the human skin from 40 to 110 ghz: A confirmation study,'' \emph{Bioelectromagnetics}, vol.~42, no.~7, pp. 562--574, 2021.

\bibitem{shu2022characterisation}
M.~Shu, X.~Shang, N.~Ridler, M.~Naftaly, C.~Guo, and A.~Zhang, ``Characterisation of dielectric materials at g-band (140--220 ghz) using a guided free-space technique,'' in \emph{2022 52nd European Microwave Conference (EuMC)}.\hskip 1em plus 0.5em minus 0.4em\relax IEEE, 2022, pp. 107--110.

\bibitem{Kazemipour2015Design}
A.~Kazemipour, M.~Hudlička, S.-K. Yee, M.~A. Salhi, D.~Allal, T.~Kleine-Ostmann, and T.~Schrader, ``Design and calibration of a compact quasi-optical system for material characterization in millimeter/submillimeter wave domain,'' \emph{IEEE Transactions on Instrumentation and Measurement}, vol.~64, no.~6, pp. 1438--1445, 2015.

\bibitem{Yashchyshyn2018A}
Y.~Yashchyshyn and K.~Godziszewski, ``A new method for dielectric characterization in sub-thz frequency range,'' \emph{IEEE Transactions on Terahertz Science and Technology}, vol.~8, no.~1, pp. 19--26, 2018.

\bibitem{Zhu2021complex}
H.-T. Zhu and K.~Wu, ``{Complex Permittivity Measurement of Dielectric Substrate in Sub-THz Range},'' \emph{IEEE Transactions on Terahertz Science and Technology}, vol.~11, no.~1, pp. 2--15, 2021.

\bibitem{zhang2023calibration}
J.~Zhang, C.~Yang, H.~Xu, and M.~Peng, ``Calibration method for terahertz free-space reflection measurement systems by imperfect short-load,'' in \emph{2023 International Conference on Microwave and Millimeter Wave Technology (ICMMT)}.\hskip 1em plus 0.5em minus 0.4em\relax IEEE, 2023, pp. 1--3.

\bibitem{Taleb2023Transmission}
F.~Taleb, G.~G. Hernandez-Cardoso, E.~Castro-Camus, and M.~Koch, ``Transmission, reflection, and scattering characterization of building materials for indoor thz communications,'' \emph{IEEE Transactions on Terahertz Science and Technology}, vol.~13, no.~5, pp. 421--430, 2023.

\bibitem{li2018compact}
L.~Li, H.~Hu, P.~Tang, R.~Li, B.~Chen, and Z.~He, ``Compact dielectric constant characterization of low-loss thin dielectric slabs with microwave reflection measurement,'' \emph{IEEE Antennas and Wireless Propagation Letters}, vol.~17, no.~4, pp. 575--578, 2018.

\bibitem{Hasar2022Complex}
U.~C. Hasar, Y.~Kaya, H.~Ozturk, M.~Izginli, J.~J. Barroso, O.~M. Ramahi, and M.~Ertugrul, ``Complex permittivity and thickness evaluation of low-loss dielectrics from uncalibrated free-space time-domain measurements,'' \emph{IEEE Transactions on Geoscience and Remote Sensing}, vol.~60, pp. 1--10, 2022.

\bibitem{hasar2022broadband}
U.~C. Hasar, Y.~Kaya, G.~Ozturk, M.~Ertugrul, and C.~Korasli, ``Broadband, stable, and noniterative dielectric constant measurement of low-loss dielectric slabs using a frequency-domain free-space method,'' \emph{IEEE Transactions on Antennas and Propagation}, vol.~70, no.~12, pp. 12\,435--12\,439, 2022.

\bibitem{Hasar2022Calibration}
U.~C. Hasar, G.~Ozturk, Y.~Kaya, and M.~Ertugrul, ``Calibration-free time-domain free-space permittivity extraction technique,'' \emph{IEEE Transactions on Antennas and Propagation}, vol.~70, no.~2, pp. 1565--1568, 2022.

\bibitem{Hasar2023dielectric}
U.~C. Hasar and H.~Ozturk, ``Dielectric constant measurement using time-domain shifted metal-backing measurements,'' \emph{IEEE Geoscience and Remote Sensing Letters}, vol.~20, pp. 1--4, 2023.

\bibitem{Guo2022Free}
Y.~Guo, Z.~Chen, Y.~Li, G.~Zhao, B.~Liang, and X.~Yao, ``Free-space measurement of dielectric and magnetic properties by double planar sample method in y-band,'' \emph{IEEE Transactions on Terahertz Science and Technology}, vol.~12, no.~2, pp. 182--192, 2022.

\bibitem{Hirata2021Measurement}
A.~Hirata, K.~Suizu, N.~Sekine, I.~Watanabe, and A.~Kasamatsu, ``Measurement of glass complex permittivity at 200-500 ghz for thz propagation simulation,'' in \emph{2020 International Symposium on Antennas and Propagation (ISAP)}, 2021, pp. 617--618.

\bibitem{Urahashi2021Complex}
M.~Urahashi and A.~Hirata, ``Complex permittivity evaluation of building materials at 200-500 ghz using thz-tds,'' in \emph{2020 International Symposium on Antennas and Propagation (ISAP)}, 2021, pp. 539--540.

\bibitem{Aliouane2022Indoor}
M.~A. Aliouane, J.-M. Conrat, J.-C. Cousin, and X.~Begaud, ``Indoor material transmission measurements between 2 ghz and 170 ghz for 6g wireless communication systems,'' in \emph{2022 16th European Conference on Antennas and Propagation (EuCAP)}, 2022, pp. 1--5.

\bibitem{olkkonen2013Complex}
M.-K. Olkkonen, V.~Mikhnev, and E.~Huuskonen-Snicker, ``{Complex permittivity of concrete in the frequency range 0.8 to 12 GHz},'' in \emph{2013 7th European Conference on Antennas and Propagation (EuCAP)}, 2013, pp. 3319--3321.

\bibitem{Abbas2001Complex}
Z.~Abbas, R.~Pollard, and R.~Kelsall, ``Complex permittivity measurements at ka-band using rectangular dielectric waveguide,'' \emph{IEEE Transactions on Instrumentation and Measurement}, vol.~50, no.~5, pp. 1334--1342, 2001.

\bibitem{Wang2020Characterization}
Y.~Wang, X.~Shang, N.~M. Ridler, T.~Huang, and W.~Wu, ``Characterization of dielectric materials at wr-15 band (50–75 ghz) using vna-based technique,'' \emph{IEEE Transactions on Instrumentation and Measurement}, vol.~69, no.~7, pp. 4930--4939, 2020.

\bibitem{GUM2008}
J.~C. for Guides~in Metrology (JCGM/WG~1), ``Evaluation of measurement data—guide to the expression of uncertainty in measurement, 1st ed., document {JCGM} 100:2008,'' \emph{Bureau International des Poids et Measures (BIPM)}, 2008.

\bibitem{Keysight2023}
\BIBentryALTinterwordspacing
K.~Technologies, ``Vector network analyzer uncertainty calculator, document {Rev. A.05.00.023.0000},'' 2023. [Online]. Available: \url{https://www.keysight.com/zz/en/lib/software-detail/computer-software/downloadable-vector-network-analyzer-uncertainty-calculator-\\1000000418epsgsud.html}
\BIBentrySTDinterwordspacing

\end{thebibliography}
\begin{IEEEbiography}[{\includegraphics[width=1in,height=1.25in,clip,keepaspectratio]{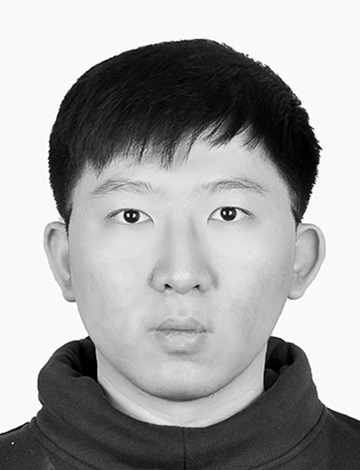}}]{Bing Xue}
(GS'23) was born in Henan, China. He is currently pursuing the D.Sc. (Tech.) degree in radio engineering in the School of Electrical Engineering, Aalto University, Espoo, Finland. His current research interests include antenna designs and measurements, human effects on antennas, and dielectric characterizations at mm-Wave frequencies for 5G systems and beyond.
\end{IEEEbiography}
\begin{IEEEbiography}[{\includegraphics[width=1in,height=1.25in,clip,keepaspectratio]{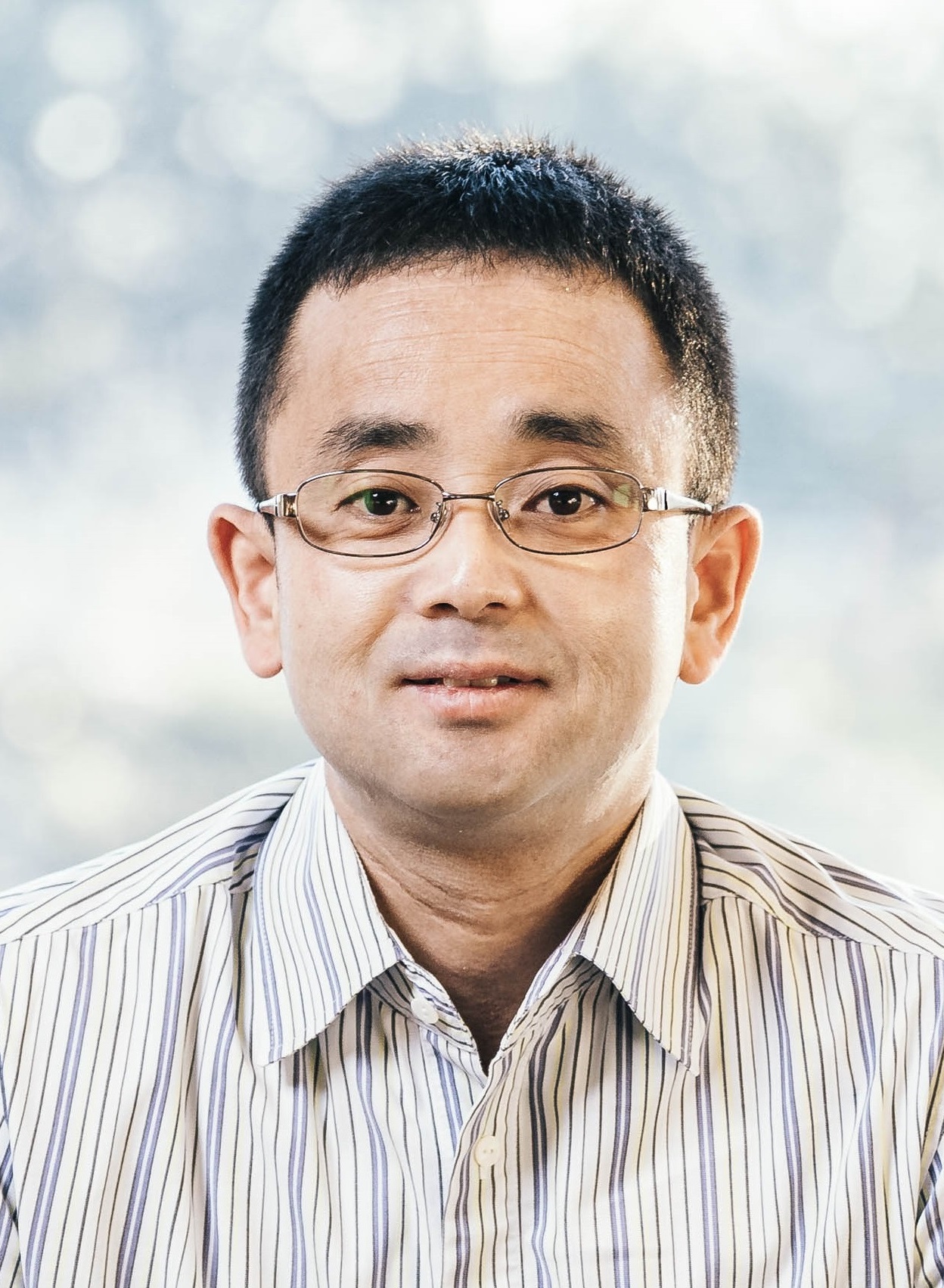}}]{Katsuyuki Haneda}
(S'03, M'07) is an associate professor in Aalto University School of Electrical Engineering, Espoo, Finland. He was an author and co-author of several best paper and student paper awards in the IEEE Vehicular Technology Conference and European Conference on Antennas and Propagation, among others. He also received the R.\ W.\ P.\ King paper award of the IEEE Transactions on Antennas and Propagation in 2021, together with Dr.\ Usman Virk. Dr. \ Haneda was an associate editor of the IEEE Transactions on Antennas and Propagation between 2012 and 2016, and of an editor of the IEEE Transactions on Wireless Communications between 2013 and 2018. He was a guest editor of Special Issues on Antennas and Propagation Aspects of In-Band Full-Duplex Applications and on Artificial Intelligence in Radio Propagation for Communications in IEEE Transactions on Antennas and Propagation in 2021 and 2022, respectively. 
Dr. Haneda was a technical programme committee co-chair of the 17th European Conference on Antennas and Propagation (EuCAP 2023), Florence, Italy. His current research activity covers high-frequency radios such as millimeter-wave and beyond and wireless for medical and smart-city applications.
\end{IEEEbiography}
\vspace{-10mm} 
\begin{IEEEbiography}[{\includegraphics[width=1in,height=1.25in,clip,keepaspectratio]{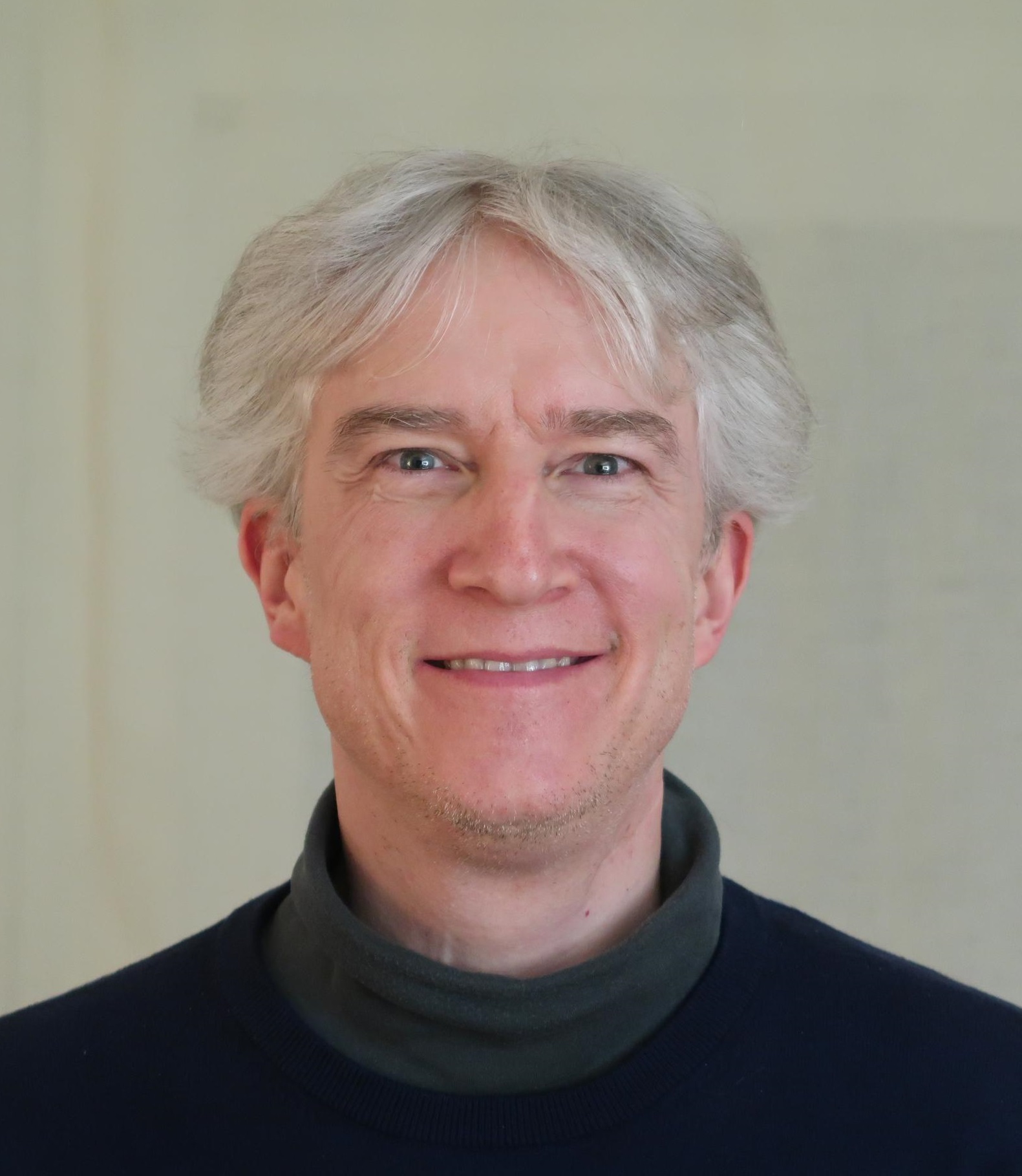}}]{Clemens Icheln}
Clemens Icheln received the Dipl.Ing. degree in electrical engineering from Hamburg-Harburg University of Technology, Hamburg, Germany, in 1996, and the Licentiate degree and the D.Sc.Tech. degree in radio engineering from Aalto University, Espoo, Finland, in 1999 and 2001, respectively. He is currently a University Lecturer with the Department of Electronics and Nanoengineering, School of Electrical Engineering, Aalto University, Espoo, Finland. His current research interests include the design of multi-element antennas for small communication devices such as mobile terminals and medical implants, to operate at frequency ranges as low as 400 MHz but also up to mm-wave frequencies, as well as the development of suitable antenna characterisation methods that allow taking, e.g., the effects of the radio channel into account. 
\end{IEEEbiography}
\vspace{-10mm} 
\begin{IEEEbiography}[{\includegraphics[width=1in,height=1.25in,clip,keepaspectratio]{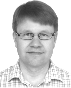}}]{Juha Ala-Laurinaho}
received the Diploma Engineer (M.Sc.) degree in mathematics and D.Sc. (Tech.) degree in electrical engineering from TKK Helsinki University of Technology, Finland, in 1995 and 2001, respectively. He has been with Aalto University, formerly TKK, serving in the Radio Laboratory in 1995–2007, in the Department of Radio Science and Engineering in 2008-2016, and at present in the Department of Electronics and Nanoengineering. Currently, he works as a Staff Scientist. Dr. Ala-Laurinaho has been a Researcher and Project Manager in various millimeter wave technology related projects. His current research interests are the antennas and antenna measurement techniques for millimeter and submillimeter waves, and the millimeter wave imaging.
\end{IEEEbiography}
\vspace{-10mm} 
\begin{IEEEbiography}[{\includegraphics[width=1in,height=1.25in,clip,keepaspectratio]{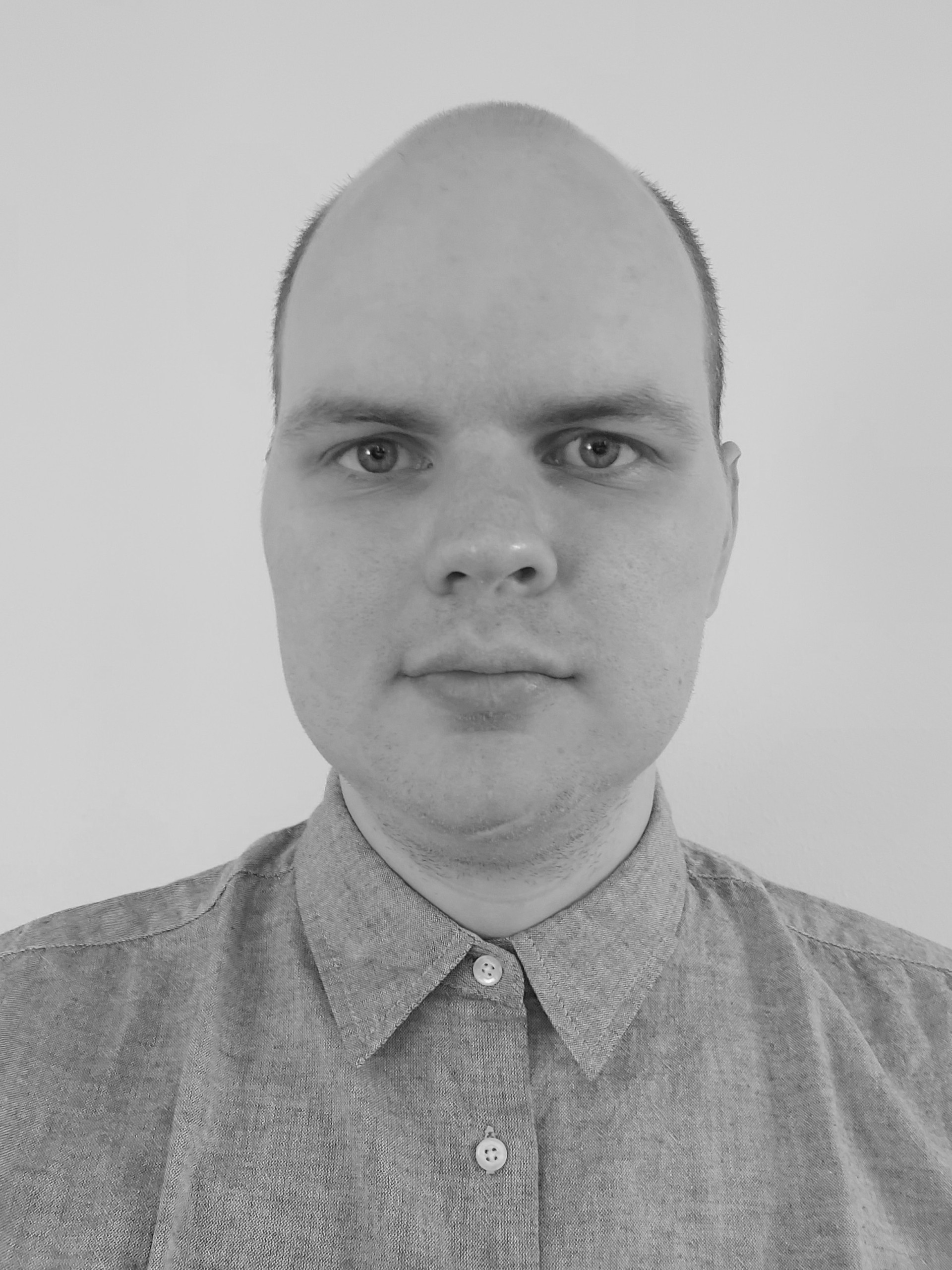}}]{Juha TUOMELA}
recieved the B.S. Degree in electrical engineering in the School of Electrical Engineering, Aalto University, Espoo, Finland in 2023 and is currently pursuing a M.S. degree in microwave engineering in Aalto University. His research work focuses on channel sounding, antenna design, and antenna measurements.
\end{IEEEbiography}
\EOD
\end{document}